\documentclass[
reprint, superscriptaddress, 
amsmath,amssymb, 
aps,prl]
{revtex4-2}

\usepackage[utf8]{inputenc}
\usepackage{graphicx}
\usepackage{dcolumn}
\usepackage{bm}
\usepackage{color}
\usepackage{enumitem}
\usepackage{amsthm}

\def\d{{\rm d}}   	
\def\R{\mathbb{R}}
\def\CP{\mathbb{CP}}

\usepackage[normalem]{ulem}

\begin{document}

\title{New asymptotically
flat Einstein–Maxwell instantons}%

\author{Bernardo Araneda}%
 \email{\texttt{baraneda@ed.ac.uk}}
 \email{\texttt{bernardo.araneda@aei.mpg.de}}
\affiliation{School of Mathematics and Maxwell Institute for Mathematical Sciences, University of Edinburgh, EH9 3FD, UK}
\affiliation{Max-Planck-Institut f\"ur Gravitationsphysik (Albert-Einstein-Institut), 
Am M\"uhlenberg 1, D-14476 Potsdam, Germany}
\author{Maciej Dunajski}%
 \email{\texttt{m.dunajski@damtp.cam.ac.uk}}
\affiliation{Department of Applied Mathematics and Theoretical Physics, University of Cambridge, Wilberforce Road, Cambridge CB3 0WA, UK}
\affiliation{Faculty of Physics, University of Warsaw Pasteura 5, 02-093 Warsaw, Poland}
\date{\today}

\begin{abstract}

We disprove the Euclidean Einstein--Maxwell Black Hole Uniqueness Conjecture, and thus demonstrate 
that the semi-classical properties of coupled gravitational and electromagnetic fields  are  more subtle than expected  from Lorentzian general relativity, where the Kerr—Newman family of metrics
yields the most general stationary and asymptotically flat black holes with a single event horizon. This is achieved by an explicit construction of a new three--parameter family of asymptotically flat Einstein--Maxwell instantons. These solutions are toric, regular, and free of conical and orbifold singularities on the manifold $M=\CP^2\setminus S^1$.  In the case of vanishing charge, these instantons reduce to the Chen--Teo Ricci flat instantons.
\end{abstract}


\maketitle

\section{Introduction}

Gravitational instantons \cite{Hawking1976} are four-dimensional, complete, non-singular, non-compact solutions to the Einstein or Einstein--Maxwell equations in Riemannian signature, with finite action, and an appropriate notion of asymptotic flatness (see e.g. \cite{Dunajski:2024pkf} for various possibilities). They provide the dominant contributions to the path integral in Euclidean Quantum Gravity (EQG) \cite{GHbook}. EQG no longer aspires to the status of a fundamental theory of quantum gravity, but
if such a theory exists, and if it reduces to general relativity as $\hbar\rightarrow 0$, then EQG
remains a valid quasi--classical limit.
It is one of few theories of quantum gravity with experimental predictions, one of which is the  derivation of the thermal properties of black holes \cite{Gibbons:1976ue} in a way that is independent from Hawking's original approach. Here, the path integral sums over all asymptotically flat (AF) metrics, which are those that near infinity approach $S^{1}\times S^{2}$. In the Ricci-flat case the known dominant contributions are the flat space $S^{1}\times\mathbb{R}^{3}$,  and the so-called `Euclidean black holes', which arise from Wick rotations of Lorentzian AF metrics. In view of the Lorentzian black hole uniqueness theorems 
(cf. \cite{Chrusciel:2012jk} for a modern review), it was natural to conjecture 
\cite{Lapedes:1980st} that all  AF  gravitational instantons belong to the two parameter family of
Euclidean Kerr solutions. This  `Euclidean Black Hole Uniqueness Conjecture' remained open for thirty years, until  2011, when an  explicit counterexample was found in a remarkable paper by Chen and Teo \cite{ChenTeo1}: there exists a 2-parameter family of AF Ricci-flat gravitational instantons that is not in the Kerr family. Therefore, given only the mass and angular momentum at infinity, one cannot determine if the solution is a Kerr or a Chen-Teo instanton, and so the semi-classical properties of the gravitational field may be more subtle than expected {\it a priori} from the mathematical structure of Lorentzian general relativity.

The instanton methods are applicable to the Einstein--Maxwell theory
(see \cite{geoff} for a recent application to charged black hole evaporations).
While Lorentzian black hole uniqueness theorems have been proven beyond the vacuum regime \cite{Chrusciel:2012jk}, the `Euclidean Black Hole Uniqueness Conjecture' does not extend to the Einstein--Maxwell case, and  Riemannian IWP metrics provide
counter examples (which, interestingly,
do  not arise as analytic continuations of black holes, as the corresponding Lorentzian solutions are
nakedly singular \cite{chruscieltod}) \cite{Dunajski:2006vs}.

It has remained an open problem, noted in \cite{ChenTeo2}
and \cite{Tod:2024msz}, to construct an Einstein-Maxwell generalisation of the Chen-Teo gravitational instanton, which would at the same time address the possible violation of the Euclidean version of the charged Lorentzian black hole uniqueness theorems. This issue was also explicitly formulated as an open problem by Yau \cite[Problem 116]{Yau}. In this note {we} settle this question,  and construct a three-parameter family of smooth, regular, AF Einstein-Maxwell instantons, which reduce to the  Chen--Teo instantons on the boundary of the parameter space. We first construct a six-parameter family of  asymptotically locally flat (ALF) Einstein-Maxwell structures:
the corresponding metrics are asymptotic to circle bundles over $S^2$, which fail to be trivial 
if the asymptotic NUT charge does not vanish. These solutions have conical singularities and generalise the vacuum metrics found in \cite{ChenTeo2}. We then demonstrate that the parameters can be adjusted to obtain a new AF Einstein-Maxwell gravitational instanton, i.e. a smooth solution free of orbifold and conical singularities and with zero asymptotic NUT charge on the manifold $M=\CP^2\setminus{S^1}$. 
We find a three-parameter family of such spaces and, in addition, show that they are conformally K\"ahler. It then follows that the Weyl tensor is one-sided type $D$, which implies that the solutions do not admit a Lorentzian section---thus there is no contradiction with Lorentzian uniqueness theorems. 

While the Ricci-flat Chen-Teo instanton was found with the inverse-scattering method \cite{ChenTeo1}, the new AF Einstein-Maxwell instanton we present was constructed using a combination of $SU(\infty)$ Toda techniques
\cite{Araneda:2023xnv,Tod:2024msz}
with the hidden symmetries of the Ernst equations \cite{MWbook}. The details will be presented in a 
separate work \cite{AD} where we also develop the general framework of toric Einstein--Maxwell metrics.

\section{ALF Einstein-Maxwell Chen-Teo geometries}

Consider a four-dimensional manifold $M$ with a metric $g$ given in local coordinates $(\tau,\phi,x,y)$ by
\begin{align}
\nonumber g ={}& \frac{(F\d\tau+G\d\phi)^2}{(x-y)FH} 
 + \frac{kH}{(x-y)^3}\left(\frac{\d{x}^2}{X} - \frac{\d{y}^2}{Y} 
 - \frac{XY\d\phi^2}{kF} \right), \\
\nonumber X ={}& a_0 + a_1 x + a_2 x^2 + a_3 x^3 + a_4 x^4, \\
\nonumber Y ={}& a_0 + a_1 y + a_2 y^2 + a_3 y^3 + a_4 y^4, \\
 F ={}& y^2 X - x^2 Y, \label{EMCT} \\
\nonumber H ={}& (\nu x+y)\left[(\nu x-y+Q(x+y))(a_1-a_3xy) \right. \\
\nonumber & \left.-(2(1-\nu)-4Q)(a_0-a_4x^2y^2) \right], \\
\nonumber G={}& [\nu^2 a_0+2\nu a_3 y^3 +(2\nu-1)a_4y^4 \\
\nonumber & -Q(\nu a_1 y + (2\nu-1)a_3 y^3 +2(\nu-1)a_4y^4) ] X \\
\nonumber & + [(1-2\nu)a_0 - 2\nu a_1x - \nu^2a_4x^4 \\
\nonumber & + Q(2(\nu-1)a_0+(2\nu-1)a_1x+\nu a_3 x^3)]Y,
\end{align}
where $a_0,...,a_4,\nu,k,Q$ are real constants. Consider also a 2-form $\mathcal{F}=\mathcal{F}^{+}+\mathcal{F}^{-}$, with
\begin{align}
\nonumber \mathcal{F}^{+} ={}& \frac{\sqrt{k}}{(\nu x+y)^2}  \left\{ 
 \d\tau\wedge({y}\d{x}-{x}\d{y}) \right. \\
 & +\frac{\d\phi}{F} \wedge [(Gy+\tfrac{HxY}{x-y})\d{x}-(Gx+\tfrac{HyX}{x-y})\d{y}]  \} \label{MaxwellField} \\
\nonumber \mathcal{F}^{-} ={}& \frac{-1}{8\sqrt{k}} 
\{\d\tau\wedge\d+\frac{\d\phi}{F}\wedge(G\d-\tfrac{H\sqrt{-XY}}{x-y}\star_{2}\d) \}\frac{H_0}{H}, 
\end{align}
where $H_0=H\big|_{Q=0}$, and $\star_2$ is the Hodge dual w.r.t the 2-metric $\frac{kH}{(x-y)^3}(\frac{\d{x}^2}{X} - \frac{\d{y}^2}{Y})$. Then, one can verify that:

\begin{enumerate}[noitemsep]
\item The pair $(g,\mathcal{F})$ is a solution to the Einstein-Maxwell equations 
with ${\mathcal F}^{\pm}=\frac{1}{2}({\mathcal F}\pm \star_g  {\mathcal F})$
\begin{align*}
& R_{ab} = 2\mathcal{F}_{ac}\mathcal{F}_{b}{}^{c}-\frac{1}{2}g_{ab}\mathcal{F}_{cd}\mathcal{F}^{cd}
\\ 
& \d  \mathcal{F}=\d \star_g \mathcal F=0.
\end{align*}
\item The geometry is ALF, and the metric $\hat{g}=\frac{(x-y)^2}{(\nu x+y)^2}g$ is K\"ahler, with K\"ahler form $\mathcal{F}^{+}$. The Weyl tensor is one-sided type $D$.
\item If $Q=0$, then \eqref{EMCT} reduces to the Ricci-flat Chen-Teo geometry \cite{ChenTeo2}.
\end{enumerate}
Two of the five coefficients $a_0, \dots ,a_4$ can be set to any non-zero values by making use of the following  symmetries of \eqref{EMCT}: $(i)$ $a_{j}\to a_{j}/c^{j}$, $(x, y, \phi)\to (c x, cy, c^2\phi)$, and $(ii)$ $a_{j}\to c a_{j}$, $\phi\to\phi/c$, where $c$ is a non-zero constant. Thus \eqref{EMCT} is a six-parameter family. 

To ensure the Riemannian signature, $x$ and $y$ must lie between two adjacent roots of the polynomial $f(\xi) = a_0 + a_1 \xi + a_2 \xi^2 + a_3 \xi^3 + a_4 \xi^4$. We shall assume that there are four different real roots $r_1,\dots,r_4$, and we take $f$ to be negative in $(r_1,r_2)$. As in \cite{ChenTeo2}, we then take
\begin{align}\label{rangesxy}
-\infty<r_1<y<r_2<x<r_3<0<r_4,
\end{align}
so the ranges of $(x,y)$ are such that $Y<0$ and $X>0$. Asymptotic infinity corresponds to $(x, y)=(r_2, r_2)$.
\begin{center}
\includegraphics[scale=0.2,angle=0]{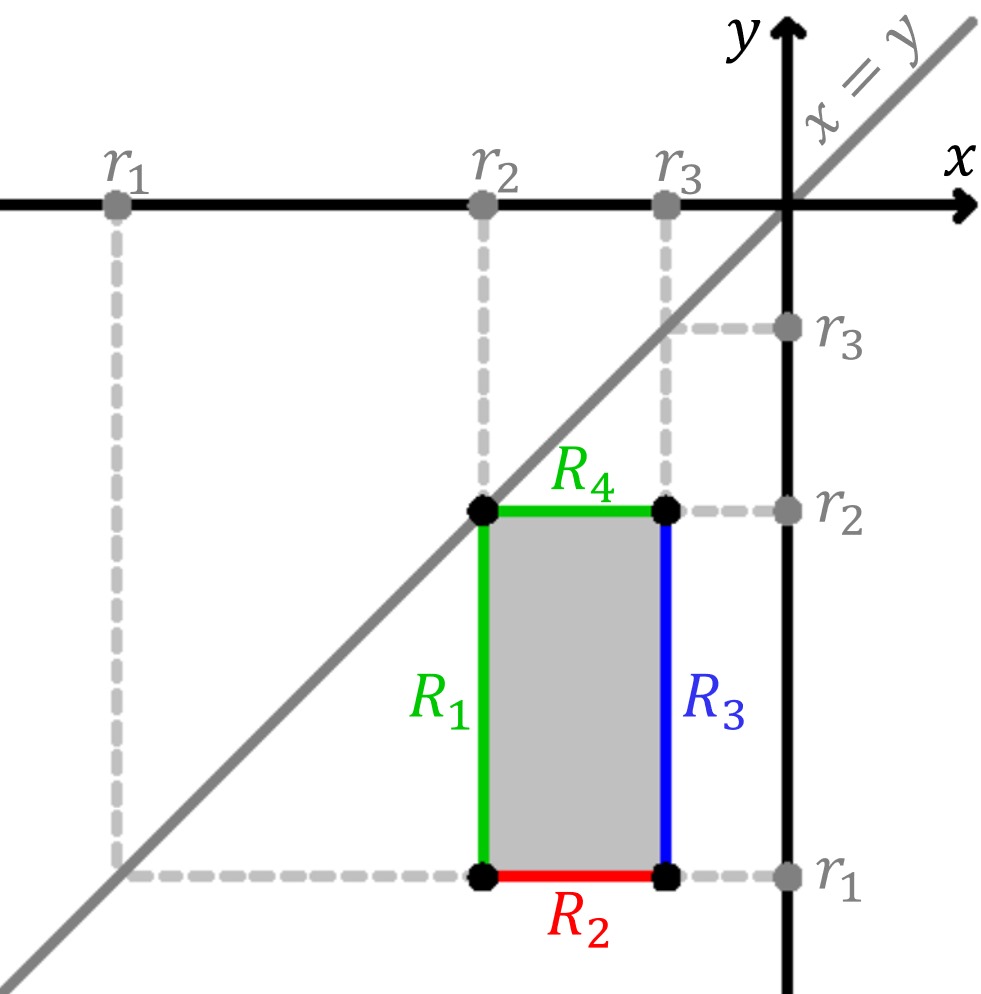}
\begin{center}
{{\bf Figure 1.} {Domain of $(x, y)$ dependence.}}
\end{center}
\end{center}

Generically, the solution has conical singularities,
and the next task is to eliminate those by restricting the values of the remaining parameters. We shall achieve it adapting  the rod structure formalism \cite{Harmark:2004rm, Chen:2010zu, Kunduri:2021xiv, Tod:2024zpa} to the Einstein--Maxwell context. There are two commuting Killing fields, $K_i\equiv \partial_{\phi^{i}}$ with $\phi^{i}=(\tau,\phi)$, generating an isometric action of $T^2$ which also preserves the Maxwell field.
The Gram matrix $J$ with components $J_{ij}=g(\partial_{\phi^{i}},\partial_{\phi^{j}})$, leads to a definition of
two coordinates $(r, z)$ given by
$
r^2=\mbox{det}(J), \; \star_2 {\d} z = {\d} r,
$
so that the metric takes a form
\[
g=\Omega^2({\d} r^2+{\d} z^2)+J_{ij}{\d}\phi^i {\d}\phi^j,
\]
where $\Omega$ and $J$ depend on $(r, z)$.
The Einstein-Maxwell equations imply that $r$ is harmonic on the orbit space, so the coordinate $z$ is indeed well defined up to adding a constant.
The space of orbits of the $T^2$ action is the half-plane $\mathbb{H}=\{(r,z),r>0\}$, with boundary $\partial\mathbb{H}=\{r=0\}$. The $T^2$ action degenerates at the boundary, where ${\rm rank}(J)<2$. Subsets of $\partial\mathbb{H}$ where ${\rm rank}(J)=1$ are called rods, and divide the $z$-axis into a series of intervals $R_1=(-\infty,z_1)$, $R_{2}=(z_1,z_2)$, \dots,  $R_{N+1}=(z_{N},\infty)$, where $z_1,\dots ,z_{N}$ are called turning points
at which $J=0$. The rods correspond to `bolts' (topological 2-spheres) or axes of rotation, while the turning points are the `nuts' \cite{Gibbons:1979xm}.
Each rod is assigned a rod vector: a Killing vector
which is a constant combination of $K_1$ and $K_2$
that vanishes on that rod.
If the Killing vector $K$ vanishes on a rod $R$, then near this rod
\[
g\sim {\Omega^{2}}|_{r=0}({\d} r^2+\kappa^2 r^2 {\d}\psi^2)+\dots
\]
where $K=\partial/\partial\psi$ and the constant $\kappa$ is the surface gravity given by
\[
\kappa=\lim_{r\rightarrow 0}\sqrt{\frac{1}{2}|\nabla K|^2}=\lim_{r\rightarrow 0} \frac{\sqrt{J(K, K)}}{\Omega r}.
\]
The metric is free of conical singularities at $r=0$ iff the coordinate $\psi$ is periodic with the period
$2\pi/\kappa$. Rescaling the Killing vector $K\rightarrow  l\equiv K/\kappa$ yields a rod direction
with a period $2\pi$ and surface gravity equal to $1$. It is convenient to represent rods by such normalised directions $l_1, \dots, l_{N+1}$ vanishing on rods $R_1, \dots, R_{N+1}$ respectively. Each of these can be expressed as $l_k={l_k}^1 K_1+{l_k}^2 K_2$ with
\[
{l_k}^i=(\Omega \sqrt{{J}_{\phi\phi}}, \Omega \sqrt{{J}_{\tau\tau}})|_{\mbox{Rod}\; k}, \quad k=1, \dots, N+1.
\]
For the solution \eqref{EMCT}, the coordinates $(r, z)$ are
\begin{align*}
& r = \frac{\sqrt{-XY}}{(x-y)^2}, \\ 
& z =\frac{2a_0+2xy(a_2+a_4 xy)+(x+y)(a_1+a_3xy)}{2(x-y)^2}.
\end{align*}
There are three turning points and four rods: $R_{1}=\{x=r_2,  r_1<y<r_2\}$, $R_{2}=\{r_2<x<r_3,  y=r_1\}$, $R_{3}=\{x=r_3,  r_1<y<r_2\}$, and $R_{4}=\{r_2<x<r_3,  y=r_2\}$. 
\begin{center}
\includegraphics[scale=0.15,angle=0]{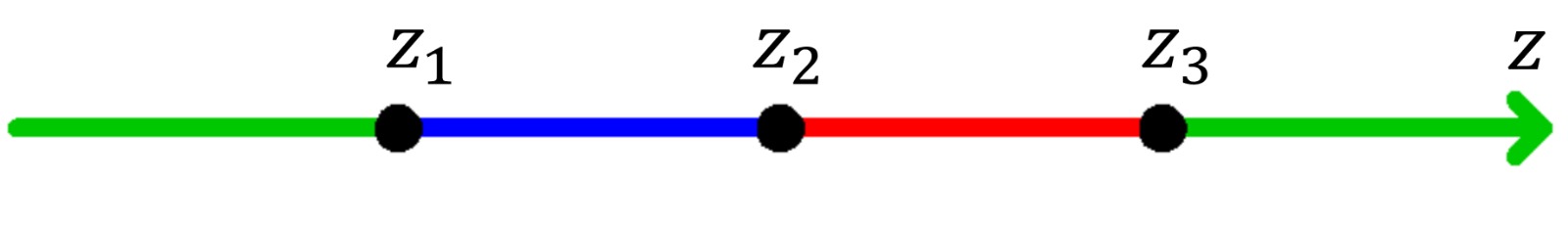}
\begin{center}
{{\bf Figure 2.} {The rod structure of (\ref{EMCT}).}}
\end{center}
\end{center}
The rod vectors are ${l}_{k}=\beta_{k}(\alpha_{k}\partial_{\tau}+\partial_{\phi})$, $k=1,\dots ,4$, where 
$\alpha_{k} = -\frac{G}{F}\big|_{R_k}$ and
\begin{align}\label{betas}
\beta_{1} = \beta_4 = \frac{2r_2\sqrt{k}}{f'(r_2)}, \;\;\;
\beta_{2} = \frac{2r_1\sqrt{k}}{f'(r_1)}, \;\;\;
\beta_{3} = \frac{2r_3\sqrt{k}}{f'(r_3)}. 
\end{align}
\section{AF Einstein--Maxwell  instantons}
We shall now show that one can adjust the parameters of the solution to get a smooth AF instanton. In order to avoid a conical singularity at the rod $R_{k}$, the orbits of the rod vector ${l}_{k}$ must be closed and have a period of $2\pi$. With \eqref{betas}, we have already taken care of this. 
To get a smooth metric we must eliminate orbifold singularities at the intersection of two adjacent rods. This is accomplished by requiring consecutive rod vectors to satisfy $({l}_{k},{l}_{k+1})=M_{k}({l}_{k-1},{l}_{k})$, where $M_{k}\in{\rm GL}(2,\mathbb{Z})$ and $\det M_{k}=\pm 1$. The solution \eqref{EMCT} has four rod vectors. The regularity condition $({l}_{2},{l}_{3})=M_{2}({l}_{1},{l}_{2})$ is equivalent to ${l}_3=-\epsilon {l}_1+p'{l}_2$, where $\epsilon=\det(M_2)=\pm1$ and $p'={\rm tr}(M_2)\in\mathbb{Z}$, and similarly $({l}_{3},{l}_{4})=M_3({l}_{2},{l}_{3})$ is equivalent to ${l}_4=-\epsilon' {l}_2+q{l}_3$, with $\epsilon'=\pm1$ and $q\in\mathbb{Z}$. 

The asymptotic NUT charge which we compute in the next section is proportional to the $\tau$--component of ${l}_1-{l}_4$. Thus, the AF condition requires ${l}_1={l}_4$ (since the $\phi$-component of ${l}_1-{l}_4$ vanishes). From the above regularity requirements, we have ${l}_1=-\epsilon{l}_3+pl_2$ (with $p=\epsilon p'$) and ${l}_4=-\epsilon'{l}_2+q {l}_3$. So ${l}_1-{l}_4=0$ gives $p=-\epsilon'$ and $q=-\epsilon$. Therefore, the combination of smoothness and asymptotic flatness is equivalent to 
\begin{align}\label{regularity}
{l}_1={l}_4=\pm({l}_2+{l}_3). 
\end{align}
We shall choose the $+$ sign, and make use of the parameter-scaling symmetry mentioned above, to set $r_{2}=-1$, $a_4=1$. The regularity condition \eqref{regularity} gives:
\begin{align}
\nonumber & r_4= 0, \\
& r_3= \frac{(1-\nu)(1+r_1)Q-\nu^2+2\nu r_1-1}{(1+r_1)((\nu-1)Q-2\nu)} \label{regroots} \\
\nonumber & r_1^2+\frac{2\nu(Q+\nu)r_1}{Q\nu-2\nu-1} +
\frac{\nu  \left(Q \nu +\nu^{2}-Q +1\right) \left(Q +\nu \right)}{\left(Q \nu -2 \nu -1\right) \left(Q \nu -Q -2 \nu \right)}=0.
\end{align}
Thus, there is a 3--dimensional moduli space of instantons parametrised by $(k, Q, \nu)$. The range of the moduli parameters are restricted by the condition \eqref{rangesxy}, which translates into $r_1<-1<r_3<0$, and by the positivity of the asymptotic norm of $\partial_{\tau}$, which is given by
\begin{align}\label{normtau}
|\partial_\tau|^{-2} \to (1+\nu)(1-\nu-2Q) >0, 
\end{align}
so that $-1<\nu<1-2Q$. The allowed region in the $(Q, \nu)$ space is bounded by the line $\nu=-1$ and a curve $\nu=(Q-2)^{-1}$. 

The four--manifold {$M$} supporting the instanton we constructed in this section is determined by the underlying rod structure. The details are as in the Ricci--flat case \cite{ChenTeo2}:  The number of turning points is equal to the Euler signature so that $\chi(M)=3$. Closing up the {two} semi–infinite rods gives the  rod structure of the complex projective plane $\CP^2$ with three
turning points as the triangle vertices, and three rods as sides. Joining the rods adds $S^1\times \R^3$ so that $M=\CP^2\setminus S^1$. 
\section{Asymptotic charges and limits}
To exhibit asymptotic flatness, we introduce a new coordinate system $(T,R,\Theta,\Phi)$ by setting 
\begin{equation}
\label{formulac}
c=\sqrt{(1+\nu)(1-\nu-2Q)}
\end{equation}
and defining 
\begin{align*}
\tau =&c \, T + \frac{1}{2}({l}_{1}^{\tau}+{l}_{4}^{\tau}  )\Phi, \quad \phi = \frac{1}{2}({l}_{1}^{\phi} +{l}_{4}^{\phi}) \Phi\\ 
x = &r_{2} - \frac{r_2 c \sqrt{k}}{R}\cos^{2}(\frac{\Theta}{2})\quad y = r_{2} + \frac{r_2 c \sqrt{k}}{R}\sin^{2}(\frac{\Theta}{2})
\end{align*}
so that $\frac{1}{2}(l_1+l_4)=\partial/\partial\Phi$.
Then, when $R\to\infty$, the metric \eqref{EMCT} 
satisfies 
\begin{align}\label{AFmetric}
 g =& (\d T+2n\cos{\Theta} \d\Phi)^2 + \d R^2\nonumber\\&+R^2(\d\Theta^2+\sin^2{\Theta}\, \d\Phi^2) + O(R^{-1}).
\end{align}
The NUT parameter $n$ can be computed from the asymptotic form of the metric \eqref{EMCT}
 as
\begin{align}
n=\frac{\sqrt{k}}{2c(r_1-r_2)(r_2-r_3)(r_3-r_4)}(n_0+n_1 Q),
\end{align}
where
\begin{align*}
n_0=&(1+\nu^2)(r_2^3+r_1r_3r_4)\\
&+2\nu r_2(r_1r_2+r_2r_3+r_2r_4 -r_1r_3-r_1r_4-r_3r_4)\\
n_1=&(1-\nu)(r_1-r_2)(r_2-r_3)(r_2-r_4).
\end{align*}
In the Ricci--flat case $Q=0$ our expression for $n$ agrees with that of Chen--Teo \cite{ChenTeo2}.
A non--zero NUT charge obstructs the asymptotic 
flatness, and the existence of globally defined Euclidean time $T$ coordinate such that the orbits of $\partial_T$ form a trivial $S^1$ bundle at infinity.

If condition (\ref{regroots}) holds then $l_4=l_1$ and $n=0$.
The leading term in (\ref{AFmetric}) with $(T, \Phi)$ 
$2\pi$--periodic is the flat metric on $S^1\times \R^3$.
We can now extract the mass $m$ and angular momentum $j$ from the sub-leading terms in \eqref{AFmetric}, i.e. from the asymptotic expansions $g_{TT}=1-\frac{2m}{R}+O(R^{-2})$ and $g_{T\Phi}=\frac{2j\sin^{2}\Theta}{R}+O(R^{-2})$ \cite{Harmark:2004rm}: we find
\begin{align}
\nonumber m ={}& \frac{1}{2}\frac{(1-\nu)}{(1+\nu)}\sqrt{k(1+\nu)(1-\nu-2Q)},\\ 
\nonumber j ={}& \frac{k((\nu-1)Q-2\nu)}{4(r_1+1)(\nu(Q-2)-1)} \{ ((r_1+3)\nu-4r_1-2)Q\\ 
&+2\nu^2+3r_1+1-2\nu(r_1+1)\}, \label{mJ}
\end{align}
where $r_1$ is given by \eqref{regroots}. When $Q=0$, these expressions agree with the mass and angular momentum for the Ricci-flat Chen-Teo instanton given in \cite{Kunduri:2021xiv}.

The Maxwell field \eqref{MaxwellField} has an asymptotic behavior
\begin{equation}
\label{Maxwellas}
\mathcal{F} =  \frac{\mathcal{Q}}{R^2}\d R\wedge \d T +\mathcal{P}\sin{\Theta}\d\Theta\wedge \d\Phi + \dots
\end{equation}
where $\mathcal{Q}$ and $\mathcal{P}$ are constants given by
\begin{equation}\label{EMcharges}
\begin{aligned}
\mathcal{Q} ={}& \lim_{R\to\infty}\frac{1}{4\pi}\int_{S_{R}}\star_{g}\mathcal{F}=
\tfrac{8k(\nu+2Q-1)+(1-\nu^2)Q}{8(1+\nu)}, \\
\mathcal{P} ={}&\lim_{R\to\infty}\frac{1}{4\pi}\int_{S_{R}}\mathcal{F}=
\tfrac{8k(\nu+2Q-1)-(1-\nu^2)Q}{8(1+\nu)}.
\end{aligned}
\end{equation}
Here, the $\dots$ terms in (\ref{Maxwellas}) correspond to the components $\mathcal{F}$, which do not contribute to the computation of
the electric and magnetic charges $\mathcal{Q},\mathcal{P}$,
and $S_{R}$ is a 2-sphere of constant $R$. 

The Maxwell field (\ref{MaxwellField}) is regular in the chosen range of $(x, y)$ coordinates, and its action is finite. To see it consider the integrand
${\mathcal F}\wedge \star_g {\mathcal F}= (|{\mathcal F}^+|^2-|{\mathcal F}^-|^2)\mbox{vol}_g$. Eliminating $r_3$ using (\ref{regroots}) we 
find, as $R\rightarrow\infty$,
\begin{eqnarray*}
|{\mathcal F}^+|^2 \mbox{vol}_g&\sim& 
\frac{2k(2Q+\nu-1)}{c(\nu Q-Q-2\nu)(\nu+1)}
 \frac{dR}{R^2}\wedge\mbox{vol}_{T^3}, \\
|{\mathcal F}^-|^2 \mbox{vol}_g&\sim& 
 -\frac{(\nu+1)(\nu-1)^2Q^2}
 {32ck(2Q+\nu-1)(\nu Q-Q-2\nu)}
 \frac{dR}{R^2}\wedge\mbox{vol}_{T^3}
\end{eqnarray*}
so that the integrals exist.

The solution (\ref{EMCT}) with $Q=0$ is the Ricci--flat five-parameter family constructed by Chen and Teo
\cite{ChenTeo2}. The corresponding Maxwell field (\ref{MaxwellField}) is  self--dual, so it does not contribute
to the energy--momentum tensor. For general $Q$
the asymptotic norm of $\partial_{\tau}$ is finite as long as $-1<\nu<1-2Q$. The boundary values $\nu=-1$ and $\nu=1-2Q$ correspond to ALE (rather than ALF or AF) metrics. If $\nu=-1$ the ASD part of the Maxwell field  vanishes and the metric $g$ belongs to the Gibbons--Hawking class with three centres in the harmonic potential lying on one axis. This is a result of a calculation where we find that
$\lim_{\nu\rightarrow -1}(H_0/H)=(1-Q)^{-1}$. The other ALE limit $\nu=1-2Q$ appears to be more subtle. If $Q=0$, then it reduces to the Pleba\'nski--Demia{\'n}ski (PD) metric \cite{ChenTeo2}. 
We expect that $\nu=1-2Q$ gives the charged PD metric, but we have not been able to verify it directly, nor to argue that the  hidden symmetry transformation we used to find
\eqref{EMCT}  reduces in the limit to a transformation relating the charged and uncharged
PD metrics.

\section{Discussion}
We  have constructed a three--parameter family of new AF Einstein--Maxwell instantons. 
 Besides its role in the path integral for quantum gravity, a physical interpretation for the new solution can be obtained via Kaluza-Klein theory. While Ricci-flat ALF solutions can be interpreted as Kaluza-Klein monopoles, the Einstein--Maxwell instantons can be lifted to solutions of five-dimensional Einstein-Maxwell theory with a Chern-Simons term, which is the bosonic sector of the lift to five dimensions of four-dimensional $N=2$ supergravity. 
 
The limiting case of our solutions are the Ricci--flat Chen--Teo instantons
 \cite{ChenTeo1, ChenTeo2} (see 
 \cite{Aksteiner:2021fae, Biquard:2021gwj, Dunajski:2024myp} for the further properties of these solutions). 
 Until very recently, {Chen-Teo} instantons provided the only counterexamples to the Euclidean Black Hole Uniqueness Conjecture. The preprint \cite{chinese} claims that the Chen--Teo solutions are just the tip of the iceberg
and there exist an AF instanton for any number of turning points in the rod structure. This, if true, may lead
to a complete classification of AF toric Ricci--flat instantons. We expect the classification of the AF toric Einstein--Maxwell instantons to be within
reach. In \cite{AD} where we present the solution generation techniques leading to (\ref{EMCT}), we shall also focus on the twistor and integrable systems 
methods \cite{MWbook,Dbook,Dunajski:2024pkf} allowing such classification.

\medskip
BA is supported by the ERC Consolidator/UKRI Frontier grant TwistorQFT EP/Z000157/1.

\end{document}